\documentclass[aps,prl,twocolumn,groupedaddress,showpacs,showkeys]{revtex4}

\usepackage{graphicx}

\begin{document}

\newcommand{\be}{\begin{equation}}
\newcommand{\ee}{\end{equation}}

\title{Hide it to see it better:\\ a robust setup to probe the thermal Casimir effect.}

\date{\today}

\author{Giuseppe Bimonte}
\affiliation{Dipartimento di Scienze
Fisiche, Universit{\`a} di Napoli Federico II, Complesso Universitario
MSA, Via Cintia, I-80126 Napoli, Italy}
\affiliation{INFN Sezione di
Napoli, I-80126 Napoli, Italy }

\begin{abstract}

We describe a  Casimir  setup consisting of two aligned sinusoidally corrugated Ni surfaces, one of which is "hidden" by a thin opaque layer of gold with a flat exposed surface.
The gold layer acts as a low-pass filter  that 
allows for a  clean observation of the controversial  thermal  Casimir force  between the corrugations, with currently available Casimir apparatuses. The proposed scheme of measurement, based on the phase-dependent modulation of the Casimir force, requires no electrostatic calibrations
of the apparatus, and is unaffected by uncertainties in the knowledge of the optical properties of the surfaces. This scheme should allow for an unambiguous discrimination
between alternative theoretical prescriptions that have been proposed in the literature for the thermal Casimir effect.

  
\end{abstract}



\maketitle

\author{ Giuseppe Bimonte}

The Casimir effect \cite{Casimir48} is the tiny force between two neutral macroscopic polarizable  bodies, that originates from quantum and thermal fluctuations of the electromagnetic (em) field in the region of space bounded by  the surfaces of the two bodies.  
Several  experiments have precisely probed the Casimir effect for surfaces made of metals, dielectrics, semiconductors, etc.. in vacuum as well as in  liquids,   
Very recently, Casimir experiments with ferromagnetic Ni  surfaces have been carried out \cite{bani1,bani2}. Numerous geometries of the surfaces have been probed, and it has been demonstrated \cite{chan,bao,chen,chiu,bani}  that the magnitude of the Casimir force can be 
controlled by  fabricating surfaces with nanoscale corrugations,  a result which opens the way to novel nanotechnological applications of the Casimir effect.  For a review see \cite{book1,book2}.

In his seminal work Casimir computed the (unit-area) force $F_C$  between two discharged perfectly conducting plane parallel plates of area $A$ at a distance $d$ in vacuum,    
obtaining the famous result:
\be
F_C=\frac{\pi^2 \hbar c}{240} \;\frac{A}{d^4}\;.
\ee     
When dealing with real surfaces, it is of course necessary to take into account a number of important corrections,
like the finite skin depth of em fields in real materials, the temperature of the plates, as well as possible imperfections of the surfaces like  roughness, absorbed impurities and especially  patch potentials  and/or stray charges on the surfaces. 

An important and yet unresolved  problem
regards the influence of the plates temperature on the magnitude of the Casimir force. 
The thermal Casimir force is quite easy to compute  for two perfectly conducting parallel plates. One finds that for separations $d$ smaller than the thermal length $\lambda_T=
\hbar c/(2 \pi k_B T)$ ($\lambda_T=1.2 \mu$m at room temperature) the thermal correction  scales like $(d/\lambda_T)^4$ and is negligibly small, while for  $d \gg \lambda_T$  it  is {\it linear} in the temperature
$F_T^{({\rm id };\infty)}=k_B T \zeta(3)/(4 \pi d^3)$ and it represents the dominant contribution to the Casimir force. It thus came as a big surprise  \cite{sernelius} when the
thermal force was computed for two plates of large but  finite conductivity (as  real metals have) because the result turned out to be  sharply different from the perfect-conductor case: the   thermal correction  was found  to  scale {\it linearly} in the temperature also  for small separations, and  it was orders of magnitude larger than the one for an ideal metal   (even though  it still represented a small correction to the total force, of a few percent for separations around 400 nm at room temperature). Moreover its limiting value for large separations   was only one-half the perfect-conductor limit $F_T^{({\rm id };\infty)}$!  
It was also found that the finite conductivity result, while consistent at large separations with the Bohr-van Leeuwen theorem of classical statistical physics \cite{Martin,bimo2},  appeared to violate Nernst third law of thermodynamics in the {\it idealized} limit of two plates with a perfect crystal structure \cite{bezerra,bezerra2}. No thermodynamic inconsistences arise however if account is taken of the finite residual resitance of real metals at low temperatures \cite{brevik2,brevik3,brevik4}. The large $T$-linear thermal correction
originates from the absence of a contribution to the Casimir force from the zero-frequency transverse-electrical TE mode (TE $\omega=0$), that follows  from the plausible assumption
(since then dubbed as Drude prescription) that 
in the limit of small frequencies the permittivity of a ohmic conductor has a $1/\omega$ singularity as in the familar Drude model. Physically, the origin of the $T$-linear correction can be traced back to 
the interaction among the Johnson currents that exist inside conductors at finite temperature \cite{john,carsten}.
A much smaller   thermal correction, proportional to $T^3$ at small distances, satisfying Nernst theorem could 
however be obtained \cite{bezerra,bezerra2} if the contribution   to the Casimir force from the  TE $\omega=0$  was calculated in accordance with the plasma model of infrared optics $\epsilon_{\rm plasma}=1-\omega_p^2/\omega^2$, with $\omega_p$ the plasma frequency  (this has since been dubbed as {\it plasma prescription}). Several small-distance Casimir experiments have been interpreted as being in accordance with the plasma prescription, and to rule out the Drude prescription (for a review of several of these experiments see \cite{critiz}). After ten years of investigations, the thermal problem is still unresolved (for a recent review see \cite{brevik}).

The view is widely shared that there is a need for experiments probing the thermal Casimir force. So far, there is only one experiment by the Yale group \cite{lamorth} which claims to have observed the thermal force between a large sphere and a plate both covered with gold, in the wide range of separations form 0.7 to 7.3 $\mu$m. The results have been interpreted by the authors as being in accordance with the  Drude prescription. This experiment has been criticized  \cite{critiz}, because the thermal Casimir force was obtained only after subtracting from the total measured force a much larger electrostatic force originating from large  patches on the surfaces.
The subtraction was perfomed by making a fit of the total observed force, based on a two-parameter model of the electrostatic force,   and not by a direct and independent measurement
of the electrostatic force, as it would have been desirable.

As of now there is no reported observation of the thermal Casimir force at separations less than 0.7 $\mu$m, which is the range in which the Casimir force has been measured most accurately. The reason is that, as note before,  the   thermal Casimir force  represents a small correction at sub micron separations,  and therefore  tight experimental demands  must be fulfilled to observe it: these include a very accurate electrostatic calibration of the apparatus  \cite{onof,onof2,onof3,decca,iannuzzi2,dalvit2},  a detailed knowledge of the optical properties of the involved surfaces  over a wide frequency range \cite{svetovoy, svetovoy2, bimonteKK} and a precise determination of the separation between the two surfaces. These factors represent sources of systematic errors that are difficult to quantify, thus complicating the interpretation  of the experiments and often stimulating  controversies. 
In this Letter we propose a novel setup  which is immune of all these complications. 
  
\begin{figure}[h]
\includegraphics[width=.9\columnwidth]{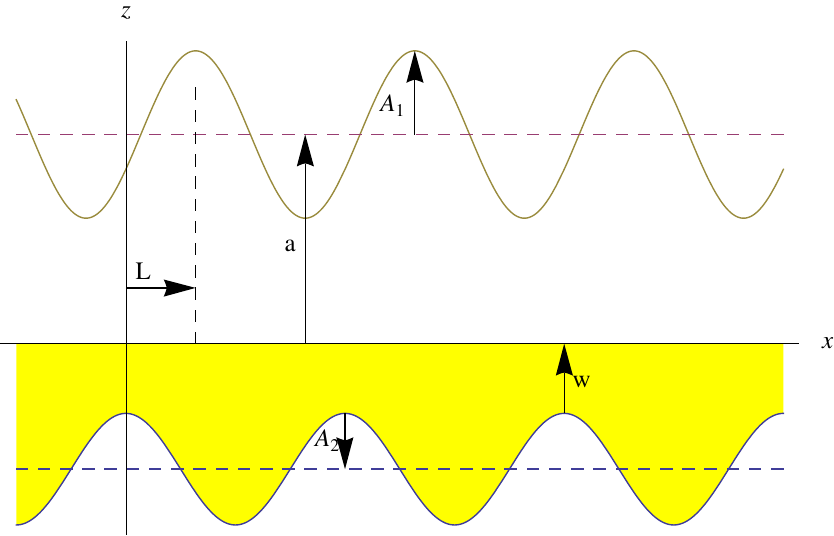}
\caption{The setup consists of two aligned uniaxial corrugated Ni plates. The lower corrugated plate is covered with a flat opaque gold layer  (shown in yellow) of
minimum thickness $w$.  }
\label{fig1}
\end{figure}

The setup, schematically shown in Fig.\ref{fig1},  consists of two  aligned uniaxially  corrugated Ni surfaces  \footnote{In a real setup the Ni surfaces would consist of Ni films with a thickness of a few hundred nanometers, as in \cite{bani1,bani2}, deposited on some substrate. As far as the Casimir force is concerned Ni films of this thickness are undistinguishable from an infinitely thick substrate.}. The  key feature  of the setup is the gold film 
with a {\it flat} exposed surface (shown in yellow),  that covers one of the two corrugations (the lower one in Fig. \ref{fig1}). The quantity to measure is  the   phase-dependent modulation  of the Casimir (unit-area) force $F(\phi)$
\be \Delta F(\phi)=F(\phi)- \frac{1}{2 \pi}\int_0^{2 \pi} d\phi F(\phi) \;,\ee where  $\phi$ the relative phase between the corrugations \footnote{The magnetic force between magnetic domains on the Ni plates, which should also show a periodic modulation, has been shown to be completely negligible in comparison to the Casimir force, for thin films of Ni like those considered in this work, in Refs. \cite{bani1,bani2}.}. 
The basic idea behind the proposed setup can be easily explained. According to the scattering approach \cite{lambr,rahi}, the Casimir force between two plates of arbitrary shapes and materials can be expressed as a discrete sum over  the so-called imaginary Matsubara  frequencies $\xi_n=2 \pi k_B T/\hbar$, $n=0,1,2,\dots$ of a series whose terms involve the product of the scattering coefficients of the two plates for em waves of frequency ${\rm i} \,\xi_n$.   The Matsubara modes that contribute significantly are those with frequencies less than a few times the characteristic frequency $\omega_c=c/(2a)$, where $a$ is the mean separation between the plates. For typical experimental separations $\omega_c$ belongs to the near UV region.  
If the minimum thickness $w$ of the gold layer in Fig.\ref{fig1} is  chosen to be much larger than the characteristic penetration depth $\delta$ of the non-vanishing Matsubara modes in gold (typically $\delta \simeq$ 10 nm or so),
we can be sure that  these  modes cannot  reach the lower Ni corrugation, 
and for them everything goes as if the lower plate were a simple flat gold surface. The conclusion of this argument, fully confirmed by detailed computations presented below, is that non-vanishing Matsubara modes give a negligible contribution to  $\Delta F$. Let us consider now the $\omega=0$ modes. They come in two types, i.e. TE and TM $\omega=0$ modes. Since TM $\omega=0$ modes represent static electric fields, they are clearly screened out by the gold overlayer and therefore their contribution to $\Delta F$ is zero.  
We are left then with the TE $\omega=0$ modes. They represent {\it static magnetic fields}, and whether they contribute or not to $\Delta F$ crucially depends on the adopted prescription. Consider first the Drude prescription.
When modeled as a Drude metal, gold is transparent to static magnetic fields.  Therefore these fields reach unimpeded the magnetic corrugation and are strongly scattered by it.  As a result, the Drude approach predicts a non-vanishing force modulation which we estimate to be measurable with presently available experimental techniques. 
Things go differently with the plasma prescription. When modeled as a dissipansionless plasma, gold effectively behaves as a superconductor and it screens out static magnetic fields within a penetration depth $\delta_0=c/\omega_p \simeq 20$ nm. Therefore, for $w \gg \delta_0$ not even the TE $\omega=0$ modes can feel the lower corrugation, and indeed we estimate that with the plasma prescription the modulation $\Delta F$ is  unmeasurably small.        
The conclusion of the above arguments is that observation of the force modulation  should allow for a clear-cut discrimination  among the two prescriptions.


The present setup presents a number of distinct advantages.  The first, and perhaps  the most important one, is that the electrostatic calibration is  unnecessary. This is a consequence of the well known fact that conductors screen out electrostaic fields. Therefore,  electric fields generated by stray electric charges and/or potential patches that may exist either on the corrugated surface of the upper Ni plate in Fig 1, or on the {\it flat} exposed surface of the gold over-layer
cannot penetrate inside the gold layer and feel the presence of the lower Ni corrugation. This implies that  the associated  electrostaic forces  are uncorrelated with the relative phase between the two corrugations, and are automatically subtracted out when we look at the modulation $\Delta F$.
  The second advantage is that the optical properties of Ni or Au are irrelevant because, as we show below, the modulation of the Casimir force  is  soley determined by the {\it static} magnetic permeability of Ni. This eliminates then two of the main sources of uncertainty that affect present measurements
of the Casimir force.

After these observations, we can present our computation of $\Delta F$. The framework for computing the Casimir force between real materials  is provided by Lifshitz theory \cite{lifs}. In its original version the material boundaries of the system are planar dielectrics ($\mu=1$)  fully described by the respective frequency-dependent  (complex) permittivity $\epsilon(\omega)$. The extension of the theory to  magneto-dielectric plates
with  non-trivial permeability $\mu(\omega)$ was worked out in \cite{rich}, and it was generalized to an arbitrary number of plane-parallel layers of magnedielectric materials  in \cite{buhm,tomas}. Recently,  Lifshitz theory has been generalized to deal with surfaces of arbitrary shapes, via  scattering methods \cite{lambr,rahi}.  For the sake of simplicity, in this Letter we shall not dwell into the complexities of scattering theory, and we shall use the simple Proximity Force Approximation (PFA) to estimate the modulation of the Casimir force for our system.  The PFA provides the Casimir force between two gently curved surfaces at close separation, and  it has been widely used to interpret  Casimir experiments \cite{book2}  (see  \cite{kruger} for more applications of the Proximity Approximation). 
Recently, it has been shown that the PFA represents  the leading term in a gradient expansion of the Casimir force, in powers of the slopes of the bounding surfaces \cite{fosco2,bimonte3,bimonte4}. The gradient expansion shows that the PFA is asymptotically exact in the zero-curvature limit, and with its help it is now possible  to estimate the error introduced by the PFA. We shall use the gradient expansion to estimate the first correction beyond PFA for our setup. 
To stay within the domain of validity of the PFA,  we shall  assume in what follows that the amplitudes $A_1$ and $A_2$ of the corrugations and the mean distance $a$ (see Fig. \ref{fig1})  are all much smaller than the period $\Lambda$.  We let $H_1(x)=a+A_1\cos[2 \pi (x-L)/ \Lambda ]$  ($0<A_1 < a$) and $H_2(x)=-w-A_2 [1-\cos(2 \pi x/\Lambda)]$ ($0<A_2$) the profiles of the two uniaxial corrugations, and we assume that the region $H_2(x) \le z \le 0$ is filled with gold.  For $A_1,A_2,\,a \ll \Lambda$, 
in the neighborhood of the point $x$ our system looks locally like a planar cavity consisting of Ni slab at distance $d(x)=H_1(x)$ from a planar two-layer   slab consisting of a layer of gold of thickness $s=-H_2(x)$ on top of a Ni substrate. If we let  $F_{\rm pp}(T,d ; s)$ the unit-area Casimir force for such a planar system, the PFA posits that the unit-area Casimir force between two gently corrugated plates is:
\be
F_{\rm PFA}=\frac{1}{\Lambda} \int_0^{\Lambda} dx F_{\rm pp}(T,H_1(x); -H_2(x))\;.\label{PFA}
\ee    According to  \cite{buhm,tomas}, the Casimir force  $F_{\rm pp}$ (per unit area) between two magnetodieletric possibly layered plane-parallel   slabs $S_{j}$, $j=1,2$ at distance $d$ in vacuum  can be represented by the formula (positive forces indicate attraction): 
$$
F_{\rm pp}(T,d)=\frac{k_B T}{\pi}\sum_{l=0}^{\infty}\left(1-\frac{1}{2}\delta_{l0}\right)\int_0^{\infty} d k_{\perp} k_{\perp} q_l
$$
\be
\times \; \sum_{\alpha={\rm TE,TM}} \left[\frac{e^{2 d q_l}}{R^{(1)}_{\alpha}({\rm i} \xi_l,k_{\perp})\;R^{(2)}_{\alpha}({\rm i} \xi_l,k_{\perp})}-1 \right]^{-1}\;,\label{lifs}
\ee
where $k_B$ is the Boltzmann constant, $\xi_l=2 \pi l k_B T/\hbar$ are the (imaginary) Matsubara frequencies, $k_{\perp}$ is the modulus of the in-plane wave-vector, $q_l=\sqrt{\xi_l^2/c^2+k_{\perp}^2}$, and $R^{(j)}_{\alpha}({\rm i} \xi_l,k_{\perp})$ is the reflection coefficient of slab $j$ for polarization $\alpha$.  The Casimir force $F_{pp}(T,d; s)$
for a planar Ni-Au-Ni system, can be  obtained from  Eq. (\ref{lifs}) by substituting  $R^{(1)}_{\alpha}$ by the Fresnel reflection coefficient $r_{\alpha}^{(0{\rm Ni})}$ of a Ni slab (given in Eqs.(\ref{freTE}) and (\ref{freTM}) below, with $a=0$, $b=$Ni),  and $R^{(2)}_{\alpha}$ by the reflection coefficient  $R_{\alpha}^{(0{\rm AuNi})}$ of a two-layer planar slab consisting of a layer of thickness $s$ of gold on a Ni substrate.  The latter reflection coefficient has the expression:

\be
R_{\alpha}^{(0{\rm Au Ni})}({\rm i} \xi_l,k_{\perp};s)=\frac{r_{\alpha}^{(0{\rm Au})}+e^{-2\,s\, k_l^{({\rm Au})}}\,r_{\alpha}^{({\rm Au Ni})}}{1+e^{-2\,s\, k_l^{({\rm Au})}}\,r_{\alpha}^{(0{\rm Au})}\,r_{\alpha}^{({\rm Au Ni})}}\;.
\ee   
Here $r^{(ab)}_{\alpha}$ are the Fresnel reflection coefficients for a planar interface separating medium a from medium b: 
\be
r^{(ab)}_{\rm TE}=\frac{\mu_{b}({\rm i} \xi_l) \,k_l^{(a)}-\mu_{a}({\rm i} \xi_l) \,k_l^{(b)}}{\mu_{b}({\rm i} \xi_l) \,k_l^{(a)}+\mu_{a}({\rm i} \xi_l) \,k_l^{(b)}}\;,\label{freTE}
\ee
\be
r^{(ab)}_{\rm TM}=\frac{\epsilon_{b}({\rm i} \xi_l) \,k_l^{(a)}-\epsilon_{a}({\rm i} \xi_l) \,k_l^{(b)}}{\epsilon_{b}({\rm i} \xi_l) \,k_l^{(a)}+\epsilon_{a}({\rm i} \xi_l) \,k_l^{(b)}}\;,\label{freTM}
\ee
where
$ k_l^{(a)}=\sqrt{\epsilon_a({\rm i} \xi_l) \mu_a({\rm i} \xi_l) \xi_l^2/c^2+k_{\perp}^2}\;$, 
 $\epsilon_a$ and $\mu_a$ denote the electric and magnetic permittivities of medium $a$, and we define $\epsilon_0=\mu_0=1$. In our computations, we modeled 
the electric permittivities  of gold  and Ni by a simple Drude model
$
\epsilon_a (\omega)=1-{\omega_a^2}/[{\omega (\omega + {\rm i}\,\gamma_a)}]
$ 
where $\omega_a$ is the plasma frequency and $\gamma_a$ is the relaxation frequency. The magnetic permittivity of Ni was modeled by the  Debye formula
$\mu(\omega)=1+({\mu(0)-1})/({1-{\rm i}\, \omega/\omega_m})\;,
$
where $\mu(0)$ is the static magnetic permeability and $\omega_m$ is a characteristic frequency. Since for typical magnetic materials $\omega_m$ is much smaller than the frequency of the first Matsubara mode $\xi_1 \sim 10^{14}$  Hz at room temperature, we can safely set $\mu=1$ in all $l>0$ terms of Eq. (\ref{lifs})   and 
just substitute the static permeability $\mu(0)=110$ \cite{bani1,bani2} in the  $l=0$ TE mode.
For the plasma frequencies and the relaxation frequencies of Au and Ni, we used the values\cite{bani1,bani2}: $\omega_{\rm Au}=9 \,{\rm eV}/\hbar$,  $\gamma_{\rm Au}=0.035\, {\rm eV}/\hbar$, $\omega_{\rm Ni}=4.89\, {\rm eV}/\hbar$ and $\gamma_{\rm Ni}=0.0436 \, {\rm eV}/\hbar$.

It is useful to separate the contribution$\Delta F_{TE}^0$ of the TE $\omega=0$   mode   from the combined contributions $\Delta {\tilde F}= \Delta F_{TM}^0+\sum_{\alpha,l > 0}\Delta F_{\alpha}^{(l) }$  of the TM $\omega=0$ plus all the non-vanishing  Matsubara modes:
$
\Delta F=\Delta F_{TE}^0+\Delta {\tilde F}\;.
$
\begin{figure}[h]
\includegraphics[width=.9\columnwidth]{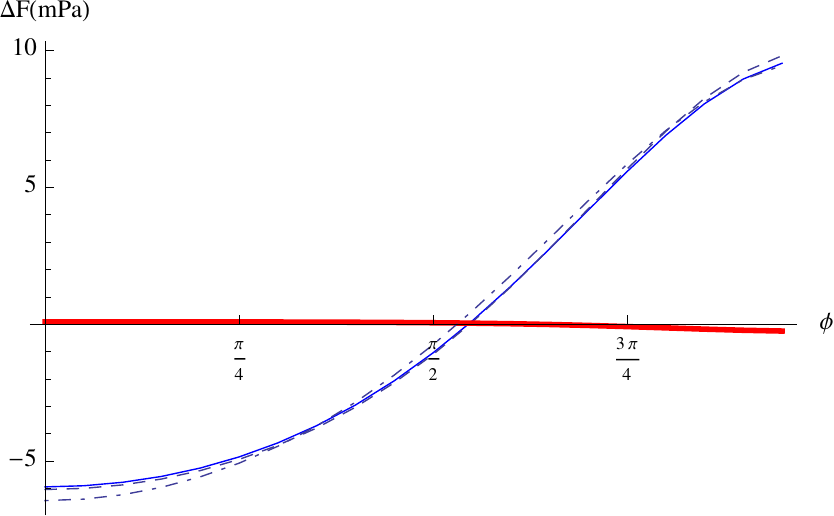}
\caption{Modulation $\Delta F$ (in mPa)  of the unit area Casimir force versus the relative phase $\phi=2 \pi L/\Lambda$ of the corrugations, computed according to the Drude prescription  (blue line). The dashed line represents the contribution $\Delta F_{TE}^0$ of the  TE $\omega=0$ mode alone. The solid red line shows the expected modulation if  the plasma prescription is used for the TE $\omega=0$ mode. The dot-dashed line includes the leading correction beyond PFA computed by the gradient expansion, for a period  $\Lambda=1\;\mu$m.}
\label{fig2}
\end{figure}
In Fig. \ref{fig2} we show (blue line) the modulation $\Delta F$ of the unit-area Casimir force (in mPa) versus the relative phase $\phi=2 \pi L/\Lambda$ of the Ni corrugations in the range $0  \le \phi \le \pi$ \footnote{It is sufficient to consider the range $0  \le \phi \le \pi$ because the modulation is symmetric under a change of sign of $\phi$.}, for $T=300$ K,  $a=140$ nm, $A_1=85$ nm, $A_2=100$ nm and $w=70$ nm.  The dashed line represents the contribution $\Delta F_{TE}^0$ of the TE zero-frequency Matsubara mode. Both the blue and the dashed lines were computed using the Drude prescription. The red line represents the modulation that obtains if the plasma prescription is used. The Figure clearly displays the main features of our setup. As it was anticipated earlier, we see that within the Drude prescription there is a periodic modulation of the Casimir force. The close proximity of the dashed and solid blue lines shows that the phase modulation of the force is entirely due to the thermal TE $\omega=0$ contribution The contribution of the non-vanishing Matsubara modes is negligible because they are screened out by the gold layer, due to their high frequencies ( $\xi_1 \sim 10^{14}$ Hz at room temperature). 
It is useful to write down the contribution $\Delta F_{TE}^0$ explicitly.
Form Eqs. (\ref{PFA}) and (\ref{lifs}) we find:
$$
F_{\rm TE \vert PFA}^{0}=\frac{k_B T}{2 \pi}  \times \frac{1}{\Lambda} \int_{0}^{\Lambda} \frac{dx}{(H_1(x)-H_2(x))^3} 
$$
\be
\times\;\int_0^{\infty}\!\!\! d y y^2
 \; \left[e^{2 y}\left(\frac{\mu(0)+1}{\mu(0)-1}\right)^2-1 \right]^{-1}\;.
\ee
We see that  $F_{\rm TE \vert PFA}^{0}$ is proportional to $T$, and that it is fully determined by the static magnetic permeability of Ni.  The modulation in Fig. \ref{fig2} is expected to be detectable. e.g. by the dynamic method used in \cite{decca2,iannuzzi3,bani1,bani2}. Within the PFA the modulation $\Delta F(\phi)$ is independent of the period $\Lambda$. 
In order to probe the reliability of the PFA, we computed the leading correction beyond PFA by using the gradient expansion \cite{fosco2,bimonte3,bimonte4}. The validity of this approach has been  demonstrated by a recent experiment with gold corrugated plates \cite{bani}.  
The resulting modulation for $\Lambda=1\;\mu$m  is shown in Fig. 2 by the dot-dashed curve.

The red curve in Fig. \ref{fig2} shows that there is essentially no modulation of the force within the plasma prescription. This is  so because within this prescription at $\omega=0$ gold effectively behaves as  a strongly diamagnetic  substance, and it expels static magnetic fields similarly to a superconductor. As a result the TE $\omega=0$ modes, that actually represent static magnetic fields get also screened out by the gold layer, and they do not feel the lower Ni corrugation.  By comparing the blue and red lines in Fig. \ref{fig2} we thus see that our setup should allow for a clean discrimination between the Drude and the plasma prescriptions for the thermal Casimir effect.

In conclusion, we have shown that a Casimir setup consisting of a pair of uniaxial  periodically corrugated aligned Ni plates one of which is covered by a thin layer of gold with a flat exposed surface, should allow for a clean observation of the thermal Casimir effect in the sub micron range, with currently available Casimir apparatuses. 
We proved that the phase modulation of the Casimir force  is a purely thermal effect, that depends only on the static magnetic permeability of the Ni plates. 
Distinctive advantages of the proposed scheme, over conventional Casimir setups, is that no electrostatic calibration of the apparatus is needed, nor any detailed 
knowledge of the optical properties of the involved materials.  An interesting  alternative to the scheme considered here consists in the replacement of Ni by a superconducting
material \cite{super1,super2}. We expect though that observation of the force modulation should be harder in this case, because for a superconductor with critical temperature $T_c$ the amplitude of the modulation would  be suppressed  by a factor of order $T_c/T_{\rm room}$ with respect to the present setup.  Another interesting  possibility would be to keep the Ni corrugation, but instead to replace the gold overlayer by a superconducting one.

\end{document}